\documentstyle[sprocl]{article}

\def\Journal#1#2#3#4{{#1} {\bf #2}, #3 (#4)}

\def\PRD{{\em Phys. Rev.} D}

\def\CQG{\em Class. Quantum Grav.}
\def\PR{Phys. Rev.}

\def\be{\begin{equation}}
\def\ee{\end{equation}}
\def\bea{\begin{eqnarray}}
\def\eea{\end{eqnarray}}

\def\be{\begin{equation}}
\def\ee{\end{equation}}
\def\bea{\begin{eqnarray}}
\def\eea{\end{eqnarray}}

\def\nn{\nonumber \\}

\def\ap{\left.}
\def\cp{\right.}
\def\at{\left(}
\def\aq{\left[}
\def\ct{\right)}
\def\cq{\right]}

\begin{document}

\title{THE EXTREMAL LIMIT OF D-DIMENSIONAL BLACK HOLES}

\author{M. CALDARELLI}

\address{Dipartimento di Fisica, via Sommarive, 
Povo,\\Trento , Italy\\E-mail: caldarel@science.unitn.it} 

\author{L. VANZO}

\address{Dipartimento di Fisica, via Sommarive, 
Povo,\\Trento , Italy\\E-mail: VANZO@science.unitn.it}

\author{S. ZERBINI}

\address{Dipartimento di Fisica, via Sommarive, 
Povo,\\Trento , Italy\\E-mail: ZERBINI@science.unitn.it}

\maketitle\abstracts{
The extreme limit of a class of D-dimensional black holes is revisited.
In the static case, It is shown that a well defined extremal limiting 
procedure exists and it 
leads to new solutions of the type 
$AdS^2 \times \Sigma_{D-2}$, $\Sigma_{D-2}$ being a 
(D-2)-dimensional constant curvature symmetric space.}

\noindent

In this contribution, we would like to revisit the extremal limit of  a large 
class of black hole solutions. The interest in the study of extreme and 
nearly-extreme black holes has recently been increased, mainly due to the 
link with one of the puzzle which is still unsolved in  the black hole 
physics: the issue related to the statistical interpretation of the 
Bekenstein-Hawking entropy.

To begin with, first  we review some particular solution of Eistein Eqs.
In classical paper, Bertotti and Robinson \cite{BR} obtained  solutions of the 
Einstein Eqs. which geometrically are the direct   product of two 2-dimensional
manifolds. If the cosmological constant is vanishing, these manifolds have 
necessarly constant curvature and opposite 
sign. Locally they are homeophorfic to $AdS_2 \times S_2$. The original 
Bertotti-Robinson solution is also globally coincident with 
$AdS_2 \times S_2$. Since these geometries are strictly related to the 
near-horizon geometry of a generic extremal black hole,  one may wonder if 
in the extremal limit, the geometry is still non trivial in the sence that 
a non vanishing Hawking temperature is present. This indeed is the case of 
De Sitter-Schwarzschild black, for which an appropiate limiting procedure
\cite{BH} leads to the Nariai solution (see \cite{BH}).    

Let us consider the Einstein Eq. with zero cosmological constant
\be
G_\mu^\nu=8\pi G T_\mu^\nu\,.
\ee
In presence of a covariant constant electric field, one may write
$\mbox{diag}T=(-E^2,-E^2,E^2,E^2)$ and we may try to solve the Einstein Eq. 
with the Bertotti-Robinson static ansaztz

\be
ds^2=-V(r)dt^2+\frac{1}{V(r)} dr^2+r^2_0 d\Omega_2^2\,,
\label{br}
\ee
where $r^2_0$ is a constant and $d\Omega_2^2$ is metric tensor of $S_2$.
A standard  elementary calculation gives
\be
G_0^0=-\frac{1}{r^2_0}\,,\,\,\,G_1^1=-\frac{1}{r^2_0}\,,
\label{e1}
\ee
\be
G_2^2=G_3^3=\frac{1}{2}V''(r)\,.
\label{e2}
\ee
Since $T_\mu^\mu=0$, one gets $G_\mu^\mu=0$. As a result
\be
-\frac{2}{r^2_0}+V''(r)=0\,.
\ee
The general solution of this elementary differential eq. may be presented in 
the form
\be
V(r)=\frac{r^2+c_1r+c_2}{r_0^2} \,,
\label{e3}
\ee
where $c_1$ and $c_2$ are two  constant. 
The Einstein Eqs. (\ref{e1}) and (\ref{e2}) are satisfied when
\be
\label{eas}
r^2_0=\frac{1}{GE^2} \,.
\ee
In summary, we have found the 2-parameter family of solutions

\be
ds^2=-(\frac{r^2+c_1r+c_2}{r_0^2})dt^2+
\frac{1}{\frac{r^2+c_1r+c_2}{r_0^2}} dr^2+r^2_0 d\Omega_2^2\,.
\label{brvz}
\ee
The original Bertotti-Robinson solution corresponds to $c_1=0$ and 
$c_2=r^2_0$\,
namely 

\be
ds^2=-(\frac{r^2}{r_0^2}+1)dt^2+
\frac{1}{(\frac{r^2}{r_0^2}+1)} dr^2+r^2_0 d\Omega_2^2\,.
\label{br0}
\ee
which is regular and globally homeomorphic to $AdS_2 \times S_2$. 

This is a consequence of the following statement, which involves the 
corresponding Euclidean sections. 

If we start from the 2-dimensional hyperbolic space,
with metric in the Poincare' form ($R$ being the radius)
\be
ds^2=\frac{R^2}{y^2}(dx^2+dy^2)\,,
\label{p}
\ee
then the mapping ($a$, $b$  suitable constant parameters)
\be
y+ix=\frac{\sqrt{r+b}-\sqrt{r+b-4aR^2}e^{-i2a \tau}}
{\sqrt{r+b}+\sqrt{r+b-4aR^2}e^{-i2a \tau}}\,,
\label{map}
\ee
reduces the Poincare' metric to
\be
ds^2=\frac{(r+b)(r+b-4aR^2)}{R^2} d\tau^2+\frac{R^2}
{(r+b)(r+b-4aR^2)}dr^2\,.
\label{pbh}
\ee

For example, choosing $R=r_0$, $b=ir_0$, $4aR^2=2ir_0$, one gets the BR 
solution, with unrestricted Euclidean time $\tau$. 

The situation changes for the  particular solution  which corresponds to the 
choice $c_1=-r^*<0$ and $c_2=0$ in Eq. (\ref{brvz}), i.e.

\be
ds^2=-(\frac{r^2-r^*r}{r_0^2})dt^2+
\frac{1}{\frac{r^2-r^*r}{r_0^2}} dr^2+r^2_0 d\Omega_2^2\,.
\label{BHvz}
\ee
This is a black hole solution with Hawking temperature

\be
\beta_H=\frac{4 \pi r_0^2}{r^*}
\ee 
This solution is only locally homeomorphic to $AdS_2 \times S_2$.
In fact, if we choose $b=0$, $R=r_0$ and $4ar_0^2=r^*$, the Euclidean 
section 
of the metric (\ref{BHvz}) reduces to the Poincare' one, but now the mapping
(\ref{map}) reads
\be
y+ix=\frac{\sqrt{r}-\sqrt{r-r^*}e^{-i2a \tau}}
{\sqrt{r}+\sqrt{r-r^*}e^{-i2a \tau}}\,.
\label{mapbh}
\ee
Here one can see that $\tau$ is defined modulo the period 
$\beta=\frac{\pi}{a}=\frac{4 \pi r_0^2}{r^*}$, which coincides with the Hawking
temperature computed requiring the absence of the conical singularity.
This may be interpreted saying that solution (\ref{BHvz}) describes a manifold 
which is only a portion of $H^2$.

A similar case is described by the choice $c_1=0$ and $c_2=-b^2 <0$, namely

\be
ds^2=-(\frac{r^2-b^2}{r_0^2})dt^2+
\frac{1}{\frac{r^2-b^2}{r_0^2}} dr^2+r^2_0 d\Omega_2^2\,.
\label{JT}
\ee
Here, the Hawking temperature computed with the standard technique is

\be
\beta_H=\frac{2 \pi r_0^2}{b}\,.
\label{cc}
\ee 
However, if we choose  $R=r_0$ and $2ar_0^2=b$, the Euclidean 
section 
of the metric (\ref{JT}) reduces to the Poincare' one, and the mapping
(\ref{map}) reads
\be
y+ix=\frac{\sqrt{r+b}-\sqrt{r-b}e^{-i2a \tau}}
{\sqrt{r+b}+\sqrt{r-b}e^{-i2a \tau}}\,.
\label{mapjt}
\ee
with $2a=\frac{b}{r^2_0}$. Again, $\tau$ is defined modulo a period
which coincides with (\ref{cc}).

It should be noted that these black hole solutions are similar to the 
Rindler space-times  and the Hawking temperature is the Unrhu one associated 
with the quantum fluctuations.

Now, let us consider a  black hole solution corresponding to a D-dimensional 
charged or neutral black hole depending on parameters as the mass $m$, charges
 $Q_i$ and the  cosmological constant $\Lambda$.  
In the Schwarzschild static coordinates (with $G=l_P^2=1$ and $ D=d+2$), it 
reads

\be
ds^2=-V(r)dt^2+
\frac{1}{V(r)} dr^2+r^2 d\Sigma_d^2\,.
\label{RN}
\ee
Here, $d\Sigma_d^2$ is the line element related to a constant curvature 
"horizon" d-dimensional manifold.
The inner and outer horizons are positive simple  roots of the shift function,
 i.e.
\be
V(r_H)=0\,,\,\,\,\,\, V'(r_H)\neq 0 \,.
\ee
The associated Hawking temperature is
\be
\beta_H=\frac{4\pi}{V'(r_H)}\,.
\ee
In general, when the extremal solution exists, namely
\be
V(r_{ex})=0\,,\,\,\,\,\, V'(r_{ex})= 0\,,\,\,\,\,\, V''(r_{ex})\neq 0 \,,
\ee
there exists a relationship between the parameters,
\be
F(m,g_i)=0\,.
\ee
When this condition is satisfied, it may happen that the original coordinates
become inappropiate (for example when $V(r)$ has a local maximum in $r=r_{ex}$,
i.e. $V''(r_{ex})<0$. 

The extreme limit has been 
investigated in several places \cite{Z,MS,MMS,NNN}.
In order to investigate the extremal limit, we introduce
the non-extremal parameter $\epsilon$ and perform the following coordinate 
change 
\be
r=r_{ex}+\epsilon r_1\,,\,\,\,\,\, t=\frac{t_1}{\epsilon}\,. 
\label{gp}
\ee
and parametrize the non-extremal limit by means of
\be
F(m,g_i)=k\epsilon^2\,,
\ee
where the sign of constant $k$ defines the physical range of the black hole
 parameters, namely the ones for which the horizon radius is non negative.
In the near-extremal limit, we may make an expansion for $\epsilon$ small.
As a consequence
\be
V(r)=V(r_{ex})+V'(r_{ex})r_1\epsilon+\frac{1}{2}
V''(r_{ex})r_1^2\epsilon^2+O(\epsilon^3)\,.
\ee
It is clear that
\be
V(r_{ex})=k_1 \epsilon^2\,,\,\,\,\,\,V'(r_{ex})=k_2\epsilon^2\,,
\ee
where $k_i$ are known constants.

Thus, the metric in the extremal limit becomes
\bea
ds^2&=&-dt_1^2 (k_1+\frac{1}{2}
V''(r_{ex})r_1^2)+\frac{dr_1^2}{(k_1+\frac{1}{2}
V''(r_{ex})r_1^2)} \nonumber \\
&+&r^2_{ex}d\Sigma_d^2\,.
\label{extrem}
\eea

As first example, let us consider the 4-dimensional charged RN black hole, 
where the horizon manifold is $S^2$ and the shift function is given by
\be
V(r)=1-\frac{2m}{r}+\frac{Q^2}{r^2}\,.
\ee
Here, the near-extremal condition reads
\be
F(m,Q)=\frac{Q^2}{m^2}-1=k\epsilon^2\,.
\ee
When $\epsilon=0$, one has $r_+=r_-=r_{ex}=m$ and $Q^2=m^2$ and the physical 
range corresponds to $k <0$, for example we may take $k=-a^2$. 
In this case, the shift function has a local minimun at $r_{ex}$ and
\be
\frac{1}{2}V''(r_{ex})=\frac{1}{m^2}\,,
\ee
but $V(r_{ex})=-a^2 \epsilon^2<0$.
as a result,
the metric in the extremal limit is
\be
ds^2=-dt_1^2 (-a^2+\frac{r_1^2}{m^2})+
\frac{dr_1^2 }{(-a^2+\frac{r_1^2)}{m^2}}
+m^2 d\Omega_2^2\,,
\label{extremRN}
\ee
and this limiting  metric describes the space-time locally $AdS_2 \times S_2$,
we have previously discussed.
This  solution 
satisfies, in the extreme limit, a BPS-like  condition, namely
\be
Q=m\,.
\ee

We note that also the Bertotti-Robinson solution may be obtained in the 
limiting procedure, but  assuming $k>0$. Thus 
 one has the local minimun for the shifth function, and this correponds to
an extremal limit within the unphysical range of black hole parameters. 

As a  second example, let us consider the Schwarzschild-DeSitter space-time.
Here the horizon manifold is still $S^2$ and since
\be
V(r)=1-\frac{2m}{r}-\frac{\Lambda r^2}{3}\,,
\ee
the solution is not saymptotically flat. As well known, there exist an  
event horizon and a cosmological horizon and
$r_H <r <r_C$.
The near extremal condition is
\be
F(m,\Lambda)=\frac{1}{3}-m\sqrt{\Lambda}=k\epsilon^2\,,\,\,\,k>0\,.
\ee
In this case, the shift function has a local maximum at the extremal 
radius $r_H=r_C=r_{ex}=(\Lambda)^{-1/2}$ and the original coordinates are 
totally inappropiate. Furthermore,  $V(r_{ex})=2k \epsilon^2 >0$
and we have
\bea
ds^2&=&-dt_1^2 (2k-\Lambda r_1^2)+\frac{dr_1^2}{2k-\Lambda r_1^2} 
\nonumber \\
&+&\frac{1}{\Lambda} d\Omega_2^2\,.
\label{extremN}
\eea
This  solution is locally $dS_2 \times S_2$ and is equivalent to the Nariai
solution \cite{BH}, a cosmological solution with $\Lambda >0$.

As further example, let us consider the 4-dimensional topological black hole
solution \cite{vanzo,brill} for which  the horizon manifold $\Sigma^2_2$ is a compact 
negative constant curvature Riemann surface and
\be
V(r)=-1-\frac{C}{r}+\frac{r^2}{l^2}\,.
\ee
Here, $l$ is related to the negative cosmological constant, namely 
$\Lambda=-\frac{3}{l^2}$ and the 
constant $C$ is given by
\be
C=m-l^* \,, \,\,\,\,l^*=\frac{2}{3\sqrt 3}l\,
\ee
$m$ being the mass of the black hole \cite{vanzo}.
The horizon radius is a positive solution of
\be
-lr^2-(m-l^*)l^2+r^3=0\,.
\ee
The extremal solution exists for $m=0$, since we have
\be
-lr^2+l^*l^2+r^3=(r-\frac{l}{\sqrt 3})^2(r+\frac{2l}{\sqrt 3})
\ee
and is given by
\be
r_{ex}=\frac{l}{\sqrt 3}\,.
\ee
As a consequence, in order to investigate the extremal limit, we may put
\be
m=C+l^*=kr_{ex} \epsilon^2 \,,\,\,\,\, k>0\,.
\ee
Thus,
\be
V(r_{ex})=-k\epsilon^2\,,\,\,\,\,k_1=-k<0\,.
\ee
and
\be
V''(r_{ex})=\frac{3}{r_{ex}^2}>0\,.
\ee
The limiting metric turns out to be
\be
ds^2=- dt_1^2[  - k+\frac{3r_1^2}{2r_{ex}^2} ]+
\frac{dr_1^2}{[ - k+\frac{3r_1^2}{2r_{ex}^2} ]}+r_{ex}^2d\Sigma_2^2\,,
\ee
namely we have $AdS_2 \times \Sigma_2$, $\Sigma_2$ being a Riemann 
surface.

Finally, we also reports  the result obtained starting from the 
Kerr-Newmann black hole solution in the standard Boyer-Linquist 
coordinates. Defining the nearly-extreme condition by means
\be
m^2-a^2-Q^2=km^2\epsilon^2\,,\,\,\,\,k>0\,,
\ee
$m$, $a$ and $Q$ being respectively the mass, the angular parameter and the 
charge of the black hole and 
making use of
\be
r=r_{ex}+\epsilon r_1\,,\,\,\,\,\, 
\phi=\phi_1+\frac{a t_1}{\epsilon(m^2+a^2)}\,, 
,\,\,\,\,\,t=\frac{t_1}{\epsilon}\,, 
\label{gps}
\ee
the limiting metric turns out to be
\bea
ds^2&=&( m^2+a^2\cos^2 \theta)\aq -\frac{r^2_1-km^2}{(m^2+a^2)^2}dt_1^2
\cp \nn
&+& \ap  \frac{dr_1^2}{r_1^2-km^2}+d\theta^2 \cq+
\frac{(m^2+a^2)^2 \sin^2 \theta}{(m^2+a^2 \cos^2 \theta)}
\at d\phi_1+\frac{2ma}{(m^2+a^2)^3}r_1dt_1 \ct^2\,. 
\eea
A similar solution (corresponding to $c=0$) has been  recently  obtained in 
\cite{bardeen}, where  one can find a detailed discussion of the related 
geometry.

In conclusion, we have shown that a large class of  black hole 
solutions
admits a well defined extremal limit procedure. In the static case, 
this procedure  gives rise to 
new solutions of the kind $AdS_2 \times \Sigma_{D-2}$,  $\Sigma_{D-2}$ being a
(D-2)-dimensional constant curvature symmetric space.

\end{document}